%% file: shearII.tex
\def\Real{{\rm I\mathchoice{\kern-0.70mm}{\kern-0.70mm}{\kern-0.65mm}%
  {\kern-0.50mm}R}}

\font \bolditalics = cmmib10
\def\bx#1{\leavevmode\thinspace\hbox{\vrule\vtop{\vbox{\hrule\kern1pt
        \hbox{\vphantom{\tt/}\thinspace{\bf#1}\thinspace}}
      \kern1pt\hrule}\vrule}\thinspace}
\def \vc #1{{\textfont1=\bolditalics \hbox{$\bf#1$}}}
\def\arcsecf {\hbox{$.\!\!^{\prime\prime}$}}
\def\arcminf {\hbox{$.\!\!^{\prime}$}}

\documentstyle[astron,12pt]{l-aa}
\input{anda9.tex}

\input psfig

\begin{document}
\thesaurus{11.05.2, 11.17.3, 12.03.2}
\title{The auto-correlation function of the extragalactic
background light: I. Measuring gravitational shear}

\author{ L. Van Waerbeke\inst{1}
\and Y. Mellier\inst{1,2,3}
\and P. Schneider\inst{2}
\and B. Fort\inst{3}
\and G. Mathez\inst{1}} 

\offprints{}
\institute{ LAT, URA285.
Observatoire Midi-Pyr\'en\'ees 14, avenue Edouard Belin 31400 Toulouse,
France \and
Max-Planck-Institut f\"ur Astrophysik, Karl-Schwarzschild-Str. 1
Postfach 1523, 85740 Garching, Germany \and
DEMIRM, Observatoire de Paris, 61 Avenue de l'Observatoire
75014 Paris, France \and 
Institut d'Astrophysique de Paris, 98 bis boulevard Arago 75014 Paris,
France}
\date{Received pretty soon, I bet ; Accepted immediately}
\maketitle

\begin{abstract}
A new method for measuring the shear induced by gravitational light
deflection is proposed. It is based on analyzing the anisotropy
induced in the auto-correlation function of the extragalactic
background light which is produced by very faint distant galaxies. The
auto-correlation function can be measured `locally', and its
anisotropy is caused by the tidal gravitational field of the
deflecting mass distribution in the foreground of these faint
background galaxies.

Since the method does not require individual galaxy detection, it
can be used to measure the shear of extremely faint galaxies which are not
detectable individually, but are present in the noise. The shear
estimated from the auto-correlation function of the noise provides an
independent measurement which can be compared to the shear obtained
from the distortion of individual galaxy images. Combining these two
independent estimates clearly increases the sensitivity of shear
measurements. In addition, our new method may allow to determine the
local magnification caused by the deflector if the auto-correlation
function is caused by a large number density of faint galaxies; in
this case, the intrinsic auto-correlation function may provide a
`standard source' with respect to which shear and magnification can be
obtained. Applications to real and synthetic data are shown and the
feasibility of our new method is demonstrated. In particular, we
present the shear maps obtained with our new method for the double QSO
2345+007 and the cluster Cl0024+16 and compare them to published shear
maps.
\end{abstract}
\section{ Introduction}
Weak gravitational lensing is one of the most promising tools for
probing the amount and the distribution of mass in the universe.  The
small anisotropy in the shape of galaxies induced by gravitational
shear maps almost directly the total mass projected near the line of
sight to these faint galaxies regardless of its dynamical state.
Recent theoretical and observational work on galaxy-galaxy lensing
(Brainerd et al., 1996, Schneider \& Rix 1996), weak lensing in rich
lensing-clusters (Kaiser \& Squires 1993, Seitz \& Schneider 1995,
Bonnet et al. 1994, hereafter BMF; Fahlmann et al. 1994, Squires et al. 1996, 
Seitz et
al. 1996), and from large scale structures (Blandford et al. 1991,
Kaiser 1992, Villumsen 1996) justifies the expectation that
outstanding results on the mass distributions of these systems will be
available in a near future which will solve numerous puzzling issues
on the (dark) mass distribution, on the number density of dark halos,
and on the relation between light and mass of gravitational
structures.

In the work quoted above, the shear is inferred from the ellipticity measured on
individual background galaxies (we shall refer to this method in the
following as `standard method', to contrast it to our new method). 
In the case of weak lensing, the induced
distortion is extremely small which makes  the measurement of the 
ellipticity from ground-based telescopes images a major concern. The 
measurement and the various intrumental and atmospheric corrections 
of shape parameters have been discussed by Kaiser, Squires \&
Broadhurst (1995)  and  
Bonnet \& Mellier (1995, hereafter, BM).  From these 
investigations, one finds that dramatic  limitations come from 
rather poor signal-to-noise ratios of images at very weak shear level. 
Ultra-deep CCD images on wide fields allow the computation of the
averaged ellipticity on a large area with many galaxies which increases
notably the signal-to-noise ratio. However, this strategy
has important shortcomings:

1. The CCD area over which ellipticities can be averaged to obtain a
local estimate of the shear is limited to a typical scale across which
the magnification matrix is approximately constant. Hence, it depends
on the typical angular scale of the gravitational
structure. Furthermore, averaging on large areas decreases the spatial
resolution of the shear map, which eventually washes out shear
produced by substructures.

2. This limitation can be overcome by ultra-deep images which provide
a large number of galaxies on small areas. Unfortunately, because
these objects are so faint, their centroid and their shape parameters
can be determined only poorly from observation.  Furthermore, deep HST
images show that a large fraction of the faintest galaxies have no
well-defined shapes and show many bright spots which can be
mistakingly considered as galaxy centroids on ground based images.

In this paper we propose an alternative method for the shear
measurement which can overcome these critical issues.  Such a method,
which uses the auto-correlation function of the brightness
distribution (hereafter ACF) of the {\sl whole} image, was implemented
by some authors (Cole et al. 1992, Dalcanton 1996) to detect small
amplitude variations in the mean noise associated with faint cluster
of galaxies.  The philosophy of the correlation methods is to avoid
individual object detection in order to work at the noise limit, and
consequently to go much deeper than classical analyses.  Here, the
fundamental point is that a stretching of a shear-free image by
gravitational distortion induces an equivalent stretching in the
auto-correlation function. The strong advantage is that the
co-addition is made without computation of the central moments of
individual objects, and, as shown later, the signal-to-noise of the
ACF can be greater than a direct co-addition of the objects.  In fact,
the concept of object itself is no more relevant: the CCD field is
viewed as a fluctuating `density field', and the objects are viewed as
`overdensities'. Consequently, any overdensity present on the CCD,
even if it is related to extremely faint sources which have not been
detected with standard detection algorithms, provides information
about the ACF. For that reason, the faintest sources, which only
emerge from the noise limit of the background, will be detected by the
ACF. It will provide additional information of the shear field from
the modification of the noise pattern which, as a result, appears as a
``{\sl sheared noise pattern}''.

Obviously, the signal-to-noise ratio of the ACF depends on the number
density of faint background sources. The most recent deep galaxy
counts in the visible at the Keck telescope (Smail et al. 1995) yields
compelling evidence that the ACF should be considerably more powerful
than the standard method. Consider for instance a standard detection
criterium based on the positive measurement of a connex-shape object
formed with at least 6 pixels at least one $\sigma$ (standard
deviation) above the background.  In practice, this definition rejects
smaller objects with flux per pixel larger than two $\sigma$ above the
background because they look like cosmic ray events (steep slope of
the contours).  Therefore, the largest ``undetected" pattern which
could be a galaxy in the CCD may have 5 pixels, having at most one
$\sigma$ above the background.  From our own observational data
obtained at the Canada-France-Hawaii Telescope (CFHT), this
``undetected'' pattern corresponds to objects with magnitude $V=30.4$
on a 5 hours exposure. Extrapolating the deep galaxy counts of Smail
et al. (1995) we expect about 2 $\times 10^6\;$gal./deg$^2$/mag. at
$V=30.4$, or about 550 gal./arcmin$^2$/mag.! In our CCD field this
amounts to about one object per field of 5 arcsec$^2$! The ACF method
will take into account any of these ultra-faint galaxies without the
need to determine their individual position and shape. We therefore
expect a considerable increase of the signal. Conversely, we may use
the increase of the signal-to-noise ratio on the shear to analyse the
projected number density of these ultra-faint sources.

In this first paper we only discuss the implication of the ACF
technique for the (weak) shear measurement. The analysis of the
background population is discussed in a forthcoming paper (Van
Waerbeke et al.; Paper II). In the next section, we present the basic
formalism and demonstrate how the ACF is ``gravitationally distorted''
by the shear. We show that the ACF method is strictly equivalent to a
single object analysis but with a high signal-to-noise ratio.  The
third section shows how to extract the gravitational shear information
from the ACF. In particular, we present a detailed analysis of the
error estimation. The fourth section validates the method on simulated
data and on real images from which we know the shear pattern from the
standard detection technique (BMF). There, we compare
shear maps obtained from the auto-correlation function with those
measured by the `standard method' for the double QSO 2345+007 and for
the inner part and an outer region of the cluster Cl 0024+16 and show
that these shear maps are in excellent agreement.

\section{ The Auto-correlation Method }
\subsection{The basic formalism}
Throughout this paper we keep solely a geometrical point of view of the
shear effects. This means that we do not consider the mass reconstruction
problem.

Consider a gravitational lens at some redshift $z_{\rm d}$ which is
sufficiently small such that the distance of all sources can be
approximately set to a constant (or more precisely, the distance ratio
$D_{\rm ds}/D_{\rm s}$ is the same for all sources). In a solid angle
which is sufficiently small such that the local lensing properties do
not change over this region, the lens equation can be locally
linearized, such that the angle $\vc\beta$ of a light ray in the
`source plane' is related to its observed angular position in the lens
plane by
\EQ
{\vc\beta}= {\vc\beta}_0 + A({\vc\theta}_0)({\vc\theta}-{\vc\theta}_0)
\EN
where $\vc\theta_0$ is a central position of the angular region
considered, $\vc\beta_0$ the corresponding angular position in the
source plane, and $A$ is the local Jacobian matrix, which reads
\EQ
A=\pmatrix{ 1-\kappa-\gamma_1 & -\gamma_2 \cr 
-\gamma_2 & 1-\kappa+\gamma_1 \cr }
\equiv (1-\kappa ){\cal I} -\gamma J(\varphi)
\EN
where $\kappa$ is the local dimensionless surface mass density, 
$\gamma_i$, $i=1,2$, are the two components of the shear (for more
details, see Schneider, Ehlers \& Falco 1992), ${\cal I}$ is the
two-dimensional identity matrix, 
and 
\EQ
J(\varphi)=\pmatrix{\cos 2\varphi & \sin 2\varphi \cr \sin 2\varphi & -\cos 2\varphi \cr}
\EN
is a symmetric traceless tensor which describes the orientation $\varphi$
of the shear. Conservation of surface brightness implies
that the observed surface brightness $I(\vc\theta)$ is related to the
intrinsic one $I^{(\rm s)}(\vc\beta)$ by
\EQ
I(\vc\theta)=I^{(\rm s)}(A \vc\theta)
\EN
where we have applied a translation of the coordinates in the source
plane to remove the constant terms in (1); this can be done since 
absolute positions in the source plane are unobservable anyway.

We now define the auto-correlation function of the brightness in the
image plane by
\EQ
\xi(\vc\theta)=\langle(I(\vc\vartheta)-\bar
I)(I(\vc\vartheta+\vc\theta)-\bar I)\rangle_\vc\vartheta
\EN
where the angular brackets denote an average over all positions
$\vc\vartheta$, and $\bar I$ is the mean surface brightness of the
image. Inserting the relation (4) into (5), one finds immediately that
\EQ
\xi(\vc\theta)=\xi^{(\rm s)}(A \vc\theta)
\EN
where the auto-correlation function $\xi^{(\rm s)}$ is defined in analogy to
(5) with $I$ replaced by $I^{(\rm s)}$. Hence we see that the `observed'
correlation function $\xi$ is related in a simple way to the
`intrinsic' one. In particular, the intrinsic correlation function can
be assumed to be isotropic; then (6) shows that the observed
correlation function will in general be anisotropic. The power
spectrum of the surface brightness, defined as the square of the
Fourier-transform $\hat\xi({\rm \bf k})$ 
of the correlation function, is related to the
intrinsic power spectrum by
\EQ
\left|\hat\xi({\rm \bf k})\right|^2=\mu^2 \left|\hat\xi^{(\rm s)}(A^{-1} {\rm \bf k})\right|^2
\EN
where $\mu=1/{\rm det} A$ is the local magnification of the lens.

If the intrinsic correlation function $\xi^{(\rm s)}(\vc\beta)$ is known, the
relation (6) allows to determine the local Jacobian matrix $A$, and
thus the shear components $\gamma$ and surface mass density
$\kappa$ (note that this is not strictly true if the lensing is not
`weak', as will be discussed later).
We shall return to the practical questions concerning this
issue further below. However, even if the intrinsic correlation
function is unknown, but only assumed to be isotropic (which should be
a very good approximation), one can obtain some information on the
Jacobian matrix $A$, as we show next.

Consider the tensor of second moments of the auto-correlation
function,
\EQ
{\cal M}_{ij}={\int {\rm d}^2 \theta\, \xi(\vc\theta)\, \theta_i\theta_j\over
\int {\rm d}^2\theta\, \xi(\vc\theta)}
\EN
and the corresponding definition for the tensor ${\cal M}_{ij}^{(\rm
s)}$ of second moments of the intrinsic auto-correlation function;
using the relation (6), one readily obtains
\EQ
{\cal M}_{ij}=A^{-1}_{ik}\, A^{-1}_{jl}\,{\cal M}_{kl}^{(\rm s)}
\EN
an equation very similar to the transformation of second brightness
moments of individual galaxy images (e.g., Blandford et al.\ts 1991,
Schneider \& Seitz 1995, BM). The main formal
difference is that individual galaxies cannot be assumed to be
isotropic in the source plane, so that the  observed ellipticity of a
galaxy image depends on both, the Jacobian matrix {\it and} the
intrinsic source ellipticity. In the case considered here,
${\cal M}_{ij}^{(\rm s)}$ can be considered an isotropic tensor, so that
\EQ
{\cal M}_{ij}^{(\rm s)}={1\over2}{\int {\rm d} \beta\,\beta^3\,\xi^{(\rm s)}(\beta) \over
\int {\rm d} \beta\,\beta\,\xi^{(\rm s)}(\beta)}\,\delta_{ij}\equiv M\,\delta_{ij}
\EN
where we have defined the mean quadratic radius $M$ of the intrinsic
correlation function in the second step. Inserting (10) into (9) and
using the form (2) of the matrix $A$, one finds after some algebra
\EQA
{\cal M}&=&{M\over (1-\kappa)^2 \left(1-\left|g\right|^2\right)^2} \nonumber\\
&&\times\pmatrix{1+2 g_1+\left|g\right|^2 & 2g_2 \cr
2 g_2 & 1-2 g_1+\left|g\right|^2 \cr }
\ENA
where we have defined the {\it reduced shear}
\EQ
g_i :={\gamma_i \over (1-\kappa)} .
\EN
We shall see that it is convenient to also introduce the {\it
distortion}
\EQ
\delta_i:={2 g_i\over (1+\left|g\right|^2)},
\EN
which has also been used in weak lensing studies before (e.g.,
Miralda-Escud\'e 1991, Schneider \& Seitz 1995, BM). With this definition, (11) can also be written as
\EQ
{\cal M}={M(1+\left|g\right|^2)\over 
(1-\kappa)^2 \left(1-\left|g\right|^2\right)^2}
\pmatrix{1+\delta_1& \delta_2 \cr
\delta_2 & 1-\delta_1 \cr } .
\EN
From the latter form, one can immediately read off that
\EQA
\delta_1&=&{{\cal M}_{11}-{\cal M}_{22}\over {\rm tr} {\cal M}} , \nonumber\\
\delta_2&=&{2{\cal M}_{12}\over {\rm tr}{\cal M}} ,
\ENA
where ${\rm tr}{\cal M}={\cal M}_{11}+{\cal M}_{22}$ is the 
trace of ${\cal M}$. Furthermore, from
the trace of (14), one obtains
\EQ
(1-\kappa)^2={M\over 4{\rm det}{\cal M}}({\rm tr}{\cal M}\pm2
\sqrt{{\rm det}{\cal M}}).
\EN
In order to appreciate this solution, the corresponding situation for
the determination of lensing parameters from individual galaxy images
should be recalled: there, using only the shape of the galaxy images,
together with the assumption that the intrinsic orientation of
galaxies is randomly distributed,
the distortion $\delta$ can be uniquely determined. Furthermore,
the local magnification can in principle 
be obtained, either from changes of the
local number density of background galaxy images due to the
magnification (anti-)bias  (Broadhurst, Taylor \&
Peacock 1995), or by changes of the image sizes at fixed surface
brightness (Bartelmann \& Narayan 1995). The same information can be
obtained in the situation under consideration here: the distortion can
be uniquely determined (15) from assuming that the intrinsic
correlation function is isotropic, without knowledge of its radial
dependence, whereas the (absolute value of the) magnification
\EQ
\left|\mu\right|={\sqrt{{\rm det} {\cal M}}\over M}
\EN
can be determined if the second moment of the intrinsic correlation $M$
function is known. Note that a given combination of $\delta$ and
$\left|\mu\right|$ does not specify the shear $\gamma$ and local surface mass
density $\kappa$ uniquely; in fact, for each such combination, four
different pairs of $(\gamma,\kappa)$ are obained, corresponding to
the four different combinations of $(\mu>0,\mu<0)$ and
$(\kappa<1,\kappa>1)$. Only one of these combinations, $\mu>0$,
$\kappa<1$, is relevant for non-critical clusters, or the region
outside the critical curves in critical clusters.

We can now consider the case of weak lensing, i.e., $\kappa\ll 1$,
$\left|\gamma\right|\ll 1$. Assuming again that the intrinsic correlation
function is isotropic, $\xi^{(\rm s)}(\vc\beta)=\xi^{(\rm s)}(\left|\vc\beta\right|)$, we
can expand (6) to obtain
\EQ
\xi(\vc\theta)\approx \xi^{(\rm s)}(\theta)-\theta\,{\xi^{(\rm s)}}^\prime(\theta) \left[\kappa+\gamma J(\varphi-\vartheta)\right] ,
\EN
where we have written $\vc\theta=\theta(\cos\vartheta,\sin\vartheta)$. 
Equation (18) shows, in the case of weak lensing, that the observed
correlation function is the sum of the intrinsic correlation function
and a perturbative term which linearly depends on both $\kappa$ and
$\gamma$.

\subsection {Inclusion of a point spread function}
Equation (4) is only an idealized description of the
brightness distribution obtained from observations; in practice, it is
affected by observational effects (seeing, tracking errors, etc.) and
noise. We consider the first of these effects here, and assume that it
can be described by a point spread function (PSF) $W(\vc\psi)$ which
is taken to be normalized,
\EQ
\int {\rm d}^2 \psi \, W({\vc\psi})=1  .
\EN
The observed brightness profile $I^{(\rm obs)}({\vc\theta})$ is then given by
\EQA
I^{(\rm obs)}(\vc\theta)&=&\int {\rm d}^2\psi\;W(\vc\theta-\vc\psi)\,I(\vc\psi)\cr
&=&\int {\rm d}^2\psi\;W(\vc\theta-\vc\psi)\,I^{(\rm s)}(A \vc\psi) .
\ENA
This observed brightness profile can now be inserted into the
definition (5) to obtain the observed correlation function
$\xi^{(\rm obs)}(\vc\theta)$, for which one obtains
\EQA
\xi^{(\rm obs)}(\vc\theta)&=&\int {\rm d}^2\psi\;T(\vc\theta-\vc\psi)\,\xi(\vc\psi)\nonumber\\
&=&\int {\rm d}^2\psi\;T(\vc\theta-\vc\psi)\,\xi^{(\rm s)}(A\vc\psi) ,
\ENA
where the self-convolved PSF $T$ is given by
\EQ
T(\vc\theta)=\int {\rm d}^2\psi\;W(\vc\theta-\vc\psi)\,W(\vc\psi) ;
\EN
as can be easily seen, the normalization of $W$ implies that $T$ is
also normalized. 

In analogy to the definition (8), one can now define the tensor
${\cal M}^{(\rm obs)}_{ij}$ of second moments of the observed auto-correlation
function; using Eq.\ts(21), one finds that
\EQ
{\cal M}^{(\rm obs)}_{ij}={\cal M}_{ij}+{\cal T}_{ij} ,
\EN
where 
\EQ
{\cal T}_{ij}:=\int  {\rm d}^2\theta\;T(\vc\theta)\,\theta_i\theta_j
\EN
is the tensor of second moments of the self-convolved PSF. Hence,
provided the PSF can be obtained with sufficient accuracy, the tensor
${\cal M}_{ij}$ can be directly obtained from (23), and thus the distortion
$\delta$ and magnification $\left|\mu\right|$ can be determined from (15) and
(17) as before. The accuracy with which ${\cal M}_{ij}$ can be obtained
depends of course on the accuracy with which $T(\vc\psi)$ is known, as
well as on the relative `size' of ${\cal M}$ and ${\cal T}$; if the components of
${\cal T}$ are much larger than those of ${\cal M}$, the determination of
${\cal M}={\cal M}^{(\rm obs)}-{\cal T}$ involves subtraction of two `big numbers' and is thus
subject to large fractional errors. 

If the PSF is isotropic, so is $T$; in this case, 
we can then write for the tensor
${\cal T}_{ij}=T\delta_{ij}$. The distortion and magnification in terms of
${\cal M}^{(\rm obs)}$ is then given by
\EQA
\delta_1&=&{{\cal M}_{11}^{(\rm obs)}-{\cal M}_{22}^{(\rm obs)} \over
{\rm tr}{\cal M}^{(\rm obs)}-2T},\nonumber\\
\delta_2&=&{2{\cal M}_{12}^{(\rm obs)} \over {\rm tr}{\cal M}^{(\rm
obs)}-2T},\nonumber\\ 
\left|\mu\right|&=&{\sqrt{{\rm det}{\cal M}^{(\rm obs)}-T {\rm
tr}{\cal M}^{(\rm obs)}+T^2}\over M} . 
\ENA

\subsection {Practical estimate of the distortion}
\subsubsection{Estimates neglecting a PSF}
The calculation of the tensor ${\cal M}$ of second moments of the correlation
function according to (8) is impossible in practice, since the
integral extends over the whole $\Real^2$, and therefore will be
dominated by noise. One would like to have the integral extend over a
finite region only. Here, we outline several possibilities how,
despite this difficulty, the distortion $\delta$ can be determined. 

Adding a weight factor $w_\theta(\left|\vc\theta\right|)$ 
in the integrals in the
definition (8),
\EQ
{\cal M}_{ij}={\int
 {\rm d}^2\theta\;\xi(\vc\theta)\,\theta_i\theta_j\,
w_\theta(\left|\vc\theta\right|) \over
\int  {\rm d}^2\theta\;\xi(\vc\theta)\,w_\theta(\left|\vc\theta\right|)},
\EN
would destroy the transformation relation (9), since this isotropic
weight function would correspond to an anisotropic weight function in
the source plane. However, the weight function
$w_\theta(\left|\vc\theta\right|)$ could be chosen to be about zero
for small values of $\left|\vc\theta\right|$ to minimize the PSF
pollution, and also for large value when $\xi$ approaches the noise
level.  In particular, the distortion can be determined without any
knowledge about the intrinsic correlation function $\xi^{(\rm s)}$,
except that it is assumed to be isotropic.  This is the method we
choose for the practical implementation of the ACF method in the next
Section.

Another possibility is to define the tensor of second moments as a
function of the Jacobian matrix; for that we note that we can write
the Jacobian as $A=a B$, where
\EQ
B=\pmatrix{1+\sqrt{1-\left|\delta\right|^2}-\delta_1 & -\delta_2 \cr
-\delta_2 &1+\sqrt{1-\left|\delta\right|^2}+\delta_1 \cr}  ,
\EN
and $a$ is a scalar. We now define the $\delta$-dependent moment
matrix
\EQ
\bar{\cal M}_{ij}={\int
 {\rm d}^2\theta\;\xi(\vc\theta)\,\theta_i\theta_j\,
w^{(\rm s)}(\left|B\vc\theta\right|) \over
\int  {\rm d}^2\theta\;\xi(\vc\theta)\,w^{(\rm s)}(\left|B\vc\theta\right|)},
\EN
where $w^{(\rm s)}(x)$ is a weight function. Note that this matrix now
explicitly depends on $\delta$, and can be calculated from the
observed correlation function for each value of $\delta$. Changing
the integration variable to $\vc\beta=A\vc\theta=a B\vc\theta$, one
finds that
\EQ
\bar{\cal M}^{(\rm s)}=A \bar{\cal M} A=a^2 B \bar{\cal M} B,
\EN
where the matrix 
\EQ
\bar{\cal M}^{(\rm s)}_{ij}={\int {\rm d}^2\beta\;\xi^{(\rm s)}(\vc\beta)\,
\beta_i\beta_j\,w^{(\rm s)}(\left|\vc\beta\right|/a)
\over \int {\rm d}^2\beta\;\xi^{(\rm s)}(\vc\beta)\,w^{(\rm s)}(\left|\vc\beta\right|/a)}
\EN
is defined in analogy to (28). What is important to note is that the
isotropy of $\xi^{(\rm s)}$ implies that $\bar{\cal M}^{(\rm s)}$ is again an isotropic
matrix. In other words, the expression
\EQ
\nu:={(\bar{\cal M}^{(\rm s)}_{11} -\bar{\cal M}^{(\rm s)}_{22})^2 +
4(\bar{\cal M}^{(\rm s)}_{12})^2
\over \left(\bar{\cal M}^{(\rm s)}_{11} +\bar{\cal M}^{(\rm s)}_{22}\right)^2}
\EN
should vanish. If $\bar{\cal M}^{(\rm s)}$ is replaced by the observed tensor
$\bar{\cal M}$ using Eq.\ts(29), $\nu$ will not vanish in general, unless
the correct value 
for $\delta$ was chosen. Performing this substitution, we find
\EQ
\nu(\delta)=1-{4{\rm det}\bar{\cal M} \left(1-\left|\delta\right|^2\right)\over
\left[{\rm tr}\bar{\cal M}-\delta_1(\bar{\cal M}_{11}-\bar{\cal M}_{22})-2\delta_2
\bar{\cal M}_{12}\right]^2}  ;
\EN
note that the scalar factor $a$ drops out.
Thus, $\nu$ depends on $\delta$ explicitly, as well as through the
dependence of $\bar{\cal M}$ on $\delta$. Since $\nu$ is positive
semi-definite, a minimization of $\nu$ will determine the value of
$\delta$. 

To summarize the basic idea of this approach, the weight function in
the definition of the second moment tensor is chosen such as to yield
an isotropic weight factor {\it in the source plane}, and it is
adjusted such that the tensor of second moments in the source plane,
as calculated from the observed tensor and the transformation (31),
becomes isotropic. The weight function $w^{(\rm s)}$ will be chosen such that it
decreases to zero at radii where the signal-to-noise of the observed
correlation function reaches about unity; for example, $w^{(\rm s)}$ could be a
Gaussian, with width adjusted to the data quality.

\subsubsection{Inclusion of a PSF}

If the effects of a PSF are included, together with a weighting
scheme, the transformation becomes much more complicated than the one
in (23). Hence, as was the case for the estimate of the distortion
from galaxy images, the analysis becomes considerably more complicated
if the effects of a PSF are taken into account (BM; Kaiser, Squires \& Broadhurst 1995). The most serious effect is
the gradual deterioration of the `isophote' shape toward the ACF center.
An issue to this problem is to use simulated images to calibrate the PSF 
effect on the shape of the objects as described in BM.
We choose this method conjointly to the biased weighting sheme (Eq.~26)
instead of the unbiased Eq. (28). The reason is that for real data, the
ACF is generally too noisy to allow a good shape parameter estimation
from the intensity weighted formula given by Eq.~(28).

Since the signal-to-noise
of the observed correlation function $\xi^{(\rm obs)}$ can be quite large,
compared to single faint galaxy images, it might be worthwile to use
an `image restoration technique' to invert the relation (21) between
the observed correlation function $\xi^{(\rm obs)}$ and the true one,
$\xi$. From the result of this restoration process (which could be,
e.g., the Lucy 1994 algorithm), the tensor $\bar{\cal M}$ can be calculated
and the method outlined above can be used. Obviously, the result will
not only depend on the quality of the data $\xi^{(\rm obs)}$, but critically
on the knowledge of the (local) PSF, or more precisely, its
self-convolved form $T$. Unfortunatly, the image restoration technique has
not given satisfactory result in the cases we have tested, and we
will not pursue this idea further here. However, we think that it should
be kept in mind and analysed in more detail.

There might be a better way for the determination of the distortion
(and the local magnification) in the presence of a PSF. The
correlation $\xi^{(\rm s)}$ is due to very many faint galaxy images if
the brighter objects are removed (see below).  Therefore, one might
hope that this intrinsic correlation function $\xi^{(\rm s)}$ is a
universal function, being the same in all directions. Such a source
correlation function will be obtainable from the Deep HST survey. If
that were true, one would have a `standard source' at high redshifts,
and it is no surprise that that would help the analysis considered
here tremendously. To wit, in this case one could compare the observed
correlation function $\xi^{(\rm obs)}$ with the one calculated from
the PSF and the intrinsic correlation function $\xi^{(\rm s)}$, i.e.,
one can define
\EQ
\chi^2=\sum_k{\left[
\xi^{(\rm obs)}(\vc\theta_k)-\int {\rm d}^2\vc\psi\;
T(\vc\theta_k-\vc\psi)\,\xi^{(\rm s)}(\left|A
\vc\psi\right|)\right]^2 \over \sigma_k^2} ,
\EN
where the sum extends over all values (pixels) on which the
correlation function $\xi^{(\rm obs)}$ is determined, and $\sigma_k$ is the
standard deviation of the value of $\xi^{(\rm obs)}$ at $\vc\theta_k$. By
minimizing $\chi^2$, the matrix $A$ can be determined. As has become
clear from the discussion at the end of Sect.\ts2.1, $\kappa$ and
$\gamma$ cannot be determined observationally, but the quantities
which can be determined are $\delta$ and $\left|\mu\right|$. Hence, for
the use in (33), one should parametrize the matrix $A$ as
\EQA
A&\doteq&{1\over \sqrt{2\left|\mu\right|\left(1-\left|\delta\right|^2
+\sqrt{1-\left|\delta\right|^2}\right)}} \nonumber\\
&\times&
\pmatrix{1+\sqrt{1-\left|\delta\right|^2}-\delta_1
& -\delta_2 \cr 
-\delta_2 &1+\sqrt{1-\left|\delta\right|^2}+\delta_1 \cr}  , 
\ENA
where the symbol `$\doteq$' implies that the right-hand side is equal
to $A$ only in the non-critical regime, but that in the critical
regime, the equivalent form of $A$ in terms of $\left|\mu\right|$ and
$\delta$ is the same as in (34). Thus, if the intrinsic correlation
function is known, one can determine $\left|\mu\right|$ and $\delta$ by
inserting the form (34) for $A$ into (33) and minimizing $\chi^2$ with
respect to $\left|\mu\right|$ and the two components of $\delta$.

\section{Practical estimation in the weak-shear limit, and error analysis}
\subsection{The practical method}
\subsubsection {Practical estimate of the correlation function}
Up to now we have considered a lens mapping which can be described by its
linearized form (1) over a sufficiently large solid angle in the
sky. In practice, if one observes a cluster of galaxies, the
linearization of the lens mapping will be valid only locally, on
scales small compared  to the typical angular scale on which the
surface mass density of the cluster varies significantly. In those
cases, one is interested in obtaining local estimates for the
distortion and the surface mass density, i.e., the average in (5)
should be extended over a limited area only.
For our task, in order to avoid boundary effects of Fourier transforms in
finite size fields, we choose to
compute the ACF from the direct method which is more CPU time
consuming than the Fourier analysis, but more accurate.
In addition, the
observational frame contains bright objects, such as foreground stars,
cluster galaxies, or bright field galaxies which would dominate the
obtained correlation function if not removed of the average in
(5). Taking these difficulties into account, we propose a practical estimator 
for the determination of the local correlation function. 

Let ${\cal U}$ denote the frame on which data are available. The bright
objects can be identified and `cut out'. This is done by using the FOCAS
package to select the bright objects and replace them by a shaded area.
It is important to note here that this shaded area does not
bias the estimate of the ACF, provided it is  not included in
the ACF computation.
The selection criteria must be such that the bright objects are
completely covered by the shade mask. The resulting frame after 
cutting out these
patches will have a `Swiss cheese' shape and will be denoted by ${\cal U}'$
afterwards. On this field ${\cal U}'$, one then defines the mean intensity
(in this section, we drop the superscript `obs' for the observed
intensity and correlation function)
\EQ
\bar{I}={1\over N'} \sum_{k\in{\cal U}'}I_k,
\EN
where the sum extends over all the $N'$ pixels in ${\cal U}'$, and the
intensity deviations $i_k=I_k-\bar I$. From that, the correlation
function of brightness fluctuations at position $\vc\theta_0$ is
defined as
\EQ
\xi(\vc\theta;\vc\theta_0)={\sum_\psi i(\vc\psi-\vc\theta/2)\,
i(\vc\psi+\vc\theta/2)\,w(\left|\vc\psi-\vc\theta_0\right|) \over
\sum_\psi w(\left|\vc\psi-\vc\theta_0\right|)} ,
\EN
where the sums in numerator and denominator are taken over all possible
values of $\vc\psi$ such that $\vc\psi\pm\vc\theta/2$ both lie in
${\cal U}'$, and $w(\theta)$ is a weight function, typically chosen to be a
Gaussian of width $\Delta \theta$. $\xi(\vc\theta;\vc\theta_0)$ is often
named Energy of the image ${\cal U}'$. Hence, the expression (36) defines
the correlation function as the weighted average over products of
pairs of points with separation vector $\vc\theta$ which can be
selected within ${\cal U}'$, and the contribution to the average is weighted
by the function $w$ which depends on the separation of the midpoint of
the pair from the point $\vc\theta_0$ at which the correlation
function is calculated.

\begin{figure}
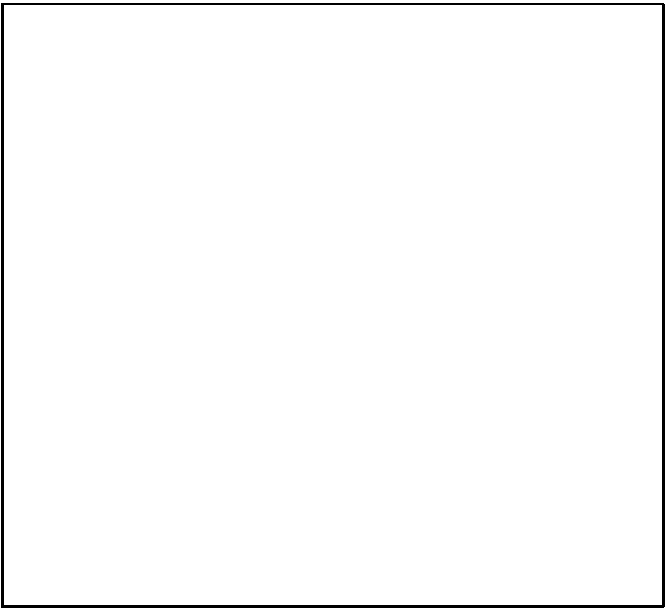

\picplace{8cm}
\caption{A characteristic auto-correlation pattern at the faint galaxy
angular scale. The ACF is computed from a $256\times 256$ superpixel of
the I band of Q2345. 
Note in the central part of the ACF 
a crux pattern. This is a possible consequence of the loss of Charge
Transfert Efficiency and/or the shift and add procedure.}
\end{figure}

An example of a correlation function is shown in Figure 1.
Only the central part of the ACF ($40\times 40$ pixels) is
represented. The rest of the image contains noise plus interference terms
which correspond to the overlapping of a galaxy by another galaxy. These
cross-correlation terms are irrelevant for our purpose.

\subsubsection{Second moment measurement}
Eq. (21) shows that the observed ACF is a convolution of the true ACF
with the square of the PSF, which is $\sqrt{2}$ times larger than the
PSF for a gaussian shape. No information is lost since the true ACF is
also $\sqrt{2}$ times larger than a typical galaxy size.  However,
this effect must be taken into account for the computation of the
moments of the ACF, in particular for the choice of the filter.  As
pointed out by BM, the usual way to estimate the second moments of an
object within a limiting isophote is not ideal. The convolution by the PSF
changes the shapes of the isophotes which become nearly circular in the
center of the object with little information left on the gravitational
shear. The best way to estimate ${\cal M}_{ij}$ seems to use an
annular weighting function $w_\theta(\vc\theta)$ centered on the ACF. Note that,
contrary to the case of individual galaxies, the center of the ACF is, by
definition, perfectly known. Since most of the gravitational shear
signal is extending between 1 and 2 seeing disks around the center of
a galaxy, a natural choice for $w_\theta$ is:
\EQ
w_\theta(\vc\theta)={\rm e}^{-\left({\theta/ s}\right)^2+1} {\rm e}^{-\left({s/ \theta}\right)^2+1}
\EN
where $s$ is the angular size of the seeing disk. 
The shear
$\gamma=\vert\gamma\vert {\rm e}^{2{\rm i}\varphi}$ 
 thus inferred must be corrected. Indeed, the magnitude
$\vert\gamma\vert$ is underestimated due to the use of the weighting
function $w_\theta$, and to the PSF convolution.

The problem of this underestimation was 
discussed earlier by several authors, as in BM, Fahlman et al. (1994),
and Kaiser, Squires \& Broadhurst (1995).
BM concluded that this effect can be
corrected using simulations from which they calibrate the measured
shear intensity to the true one. They showed that the measured
shear intensity is simply proportional to the true one (provided that
the weak shear approximation applies). Fahlman et al. showed that the
calibration constant can be found from a streching 
of the images and convolving them with a known and large isotropic PSF.
Kaiser et al. proposed to use HST images, stretch them by a known
shear, reconvolve them with a PSF typical for ground-based
observations, and then compare the resulting value of the measured
`shear' with the input value to obtain the correction factor between
true and observed shear.

As in BM, we choose here to use simulated images for the calibration.
However, HST images provide an
elegant way to calibrate the effect, as quoted in Section 2.3.2, but
this requires a deep ``empty field'', and will be done in further
work.

\subsection{ Errors coming from the noise level of the ACF}

This Section is devoted to the error measurement coming from the noisy
aspect of the ACF. This is an important issue: contrary to standard methods 
where the shear is
inferred from the averaged shapes of many galaxies, the ACF method
gives ultimately the shape of only one object, the ACF itself. It is 
therefore crucial to properly quantify the various sources of noise.

In the first subsection, the signal-to-noise ratio of the ACF ($S/N$)
is expressed in terms of the galaxy number $N_{\rm g}$, the field
size, and the $S/N$ of the individual galaxies, called $\sigma_0$.
The second subsection quantifies the dependence of the shear
measurement on the $S/N$ of the ACF.

\subsubsection{Dependence of the $S/N$ of the ACF on the image
parameters}
To simplify the discussion, assume that the image contains $N'_{\rm
g}$ galaxies having the same size, the same profile, and the same
surface brightness. We can then write the surface brightness as
$I(\vc\theta)=I_{\rm sky}+i_{\rm n}(\vc\theta)+\sum_{k=1}^{N'_{\rm g}}
I_{\rm g}(\vc\theta-\vc\theta_k)$, where $I_{\rm sky}$ is the sky
brightness, $i_{\rm n}(\vc \theta)$ are the brightness fluctuations,
and the third term is the contribution from the galaxies, situated at
positions $\vc\theta_k$, with brightness profile $I_{\rm
g}$. Subtracting the mean $\bar I$ from $I(\vc\theta)$, one obtains
that $i(\vc\theta)=
\sum_{k=1}^{N'_{\rm g}} I_{\rm g}(\vc\theta-\vc\theta_k)-\bar I_{\rm g}
+i_{\rm n}(\vc\theta)\equiv \sum_{k=1}^{N'_{\rm g}} \hat I_{\rm
g}(\vc\theta-\vc\theta_k)+ i_{\rm n}(\vc\theta)$, where $\bar I_{\rm
g}$ is the mean brightness contributed by the galaxies. Since the
galaxies are faint, and we are far from the confusion limit,
practically all photons can be attributed to the sky brightness, which
implies that the noise $i_{\rm n}(\vc\theta)$ is Poisson noise almost
entirely due to the sky brightness photons, i.e., $\sigma_{\rm
sky}^2\equiv\langle i_{\rm n}^2(\vc\theta)\rangle =I_{\rm sky}$, if
intensities are measured in photon numbers. In particular, this means
that the noise is uncorrelated with the signal.

Inserting the foregoing expression for $i(\vc\theta)$ into (36), one
finds that
\EQA
\xi(\vc\theta;\vc\theta_0)&=&
{1\over \sum_\psi w(\left|\vc\psi-\vc\theta_0\right|)}
\sum_{\psi} w(\left|\vc\psi-\vc\theta_0\right|) \cr
&\times& \Big[i_{\rm n} (\vc\psi-\vc\theta/2) \, i_{\rm n}
(\vc\psi+\vc\theta/2) \cr
&+&i_{\rm n} (\vc\psi-\vc\theta/2)\sum_{k=1}^{N'_{\rm g}}\hat I_{\rm
g}(\vc\psi+\vc\theta/2 -\vc\theta_k) \cr
&+&i_{\rm n} (\vc\psi+\vc\theta/2)\sum_{k=1}^{N'_{\rm g}}\hat I_{\rm
g}(\vc\psi-\vc\theta/2 -\vc\theta_k) \cr
&+&\sum_{j=1}^{N'_{\rm g}}\hat I_{\rm
g}(\vc\psi-\vc\theta/2 -\vc\theta_j)
\sum_{k=1}^{N'_{\rm g}}\hat I_{\rm
g}(\vc\psi+\vc\theta/2 -\vc\theta_k) \Big]
\ENA
One thus sees that $\xi$ contains contributions from galaxies only,
from products of galaxy brightness and the noise, and terms
proportional to the square of the noise. Considering $\vc\theta\ne \vc
0$, then the fact that the noise in different pixels is uncorrelated,
and that the noise is uncorrelated with the signal contributed by the
galaxies implies that the expectation value of $\xi$ is determined by
the final term in (38). This can be split up into terms with $j=k$,
and in those with $j\ne k$. The former thus contain the ACF of the
galaxy brightness profiles and its result is proportional to $N'_{\rm
g}$. The latter contains cross-correlations of different galaxies, and
is proportional to ${N'_{\rm g}}^2$ times the two-point angular
correlation function of galaxy positions.

More precisely, if we define the auto-correlation function of an
individual galaxy as 
\EQ
\xi_{\rm g}(\vc\theta)=\int{\rm d}^2\vartheta\;\hat I_{\rm
g}(\vc\vartheta-\vc\theta/2)\, \hat I_{\rm
g}(\vc\vartheta+\vc\theta/2)\; ,
\EN
and neglect the possible angular correlation of galaxy positions
[i.e., the terms with $j\ne k$ in the last term in (38)], then one
finds that the expectation value $\langle
\xi(\vc\theta;\vc\theta_0)\rangle$ of the ACF becomes
\EQ
\langle\xi(\vc\theta;\vc\theta_0)\rangle
={\sum_{j=1}^{N'_{\rm g}}w\left(\vert
\vc\theta_j-\vc\theta_0\vert\right) \over 
\sum_\psi w\left(\vert \vc\psi-\vc\theta_0\vert\right) }
\,\xi_{\rm g}(\vc\theta)\; ,
\EN
provided the weight function $w(\theta)$ does not vary on scales
comparable to the size of individual galaxies. Correspondingly, the
noise in the ACF becomes
\EQ
\sqrt{\left\langle\left[\xi-\langle\xi(\vc\theta;\vc\theta_0)
\rangle\right]^2\right\rangle}=\sigma_{\rm sky}^2
{\sqrt{\sum_\psi w^2\left(\vert \vc\psi-\vc\theta_0\vert\right) }\over
\sum_\psi w\left(\vert \vc\psi-\vc\theta_0\vert\right)}\; .
\EN
For the special case of a rectangular top-hat filter with $N_x\times
N_y$ pixels, centered on $\vc\theta_0$, the corresponding
signal-to-noise ratio of the ACF becomes
\EQ
S/N={N_{\rm g} \over \sqrt{N_x\,N_y}} {\xi_{\rm g}(\theta)\over
\sigma_{\rm sky}^2} \; ,
\EN
where now $N_{\rm g}$ is the number of galaxies contained in the
top-hat filter. We thus see that the signal-to-noise ratio is
proportional to the number of galaxies within the filter. This should
be contrasted to the case where the shear is determined from
individual galaxy images; in that case, the signal-to-noise increases
only as the square root of $N_{\rm g}$. This fact shows that the ACF
method will be superiour to the standard method if the number
(density) of galaxies is sufficiently high. Furthermore, $S/N$ depends
on the filter scale, or in other words, depends on the number density $n$
of the galaxies, $S/N\propto n\, \sqrt{N_x\,N_y}$.

The $S/N$ also depends on the flux of the individual galaxies. This is
a consequence of the fact that, by definition, the objects in the ACF
are weighted by the square of their flux [see Eq. (36)].  Therefore
the $S/N$ increases as the square of the flux of the individual
galaxies; that is as $\sigma_0^2$. 

In this simple discussion where we neglected the profiles and the sizes
of individual galaxies, we can therefore express $S/N$ as follows:
\EQ
{S\over N}= K \sqrt{{256\times 256\over N_x N_y}} N_{\rm g} \sigma_0^2\;.
\EN
The constant $K$ can be determined from the profile and
size of individual galaxies; here, we obtain it from
the hypothesis that all the galaxies have the same size, profile and
flux. They are modeled by a gaussian profile of diameter 5 pixels.
We define the $S/N$ of an object (ACF or galaxy) as
the mean amplitude of the object in an annulus of radius 3 pixels with
a width of 1 pixel centered on the object, divided by the sky photon
noise.  As explained in Section 3.1.2, the inner and outer radii of
the annulus are justified by the fact that we avoid the central part
of the object which is strongly polluted by the PSF.

With these definitions, and for a square-shaped two-dimensional top-hat
filter $w$, we find that the $S/N$ becomes
\EQ
{S\over N}=35 \  \sqrt{{256\times 256\over N_x N_y}} 
\left({N_{\rm g} \over 600} \right) \left({ \sigma_0 \over 1}\right)^2
\EN
This law was checked on a large number of numerical simulations.

For large $N_{\rm g}$, the linear dependence of $S/N$ with $N_{\rm g}$
is a strong gain of the ACF method compared to the standard method of
coadding images method (BM) for which the signal-to-noise increases as
$\sqrt{N_{\rm g}}$.  A comparison of the $S/N$ for both the ACF and BM
method is shown on Figure 2. In the case of deep observations (large
$N_{\rm g}$), the ACF will provide an easier and more accurate shear
estimate from the shape matrix of the ACF than the coadded images of
galaxies.

\begin{figure}
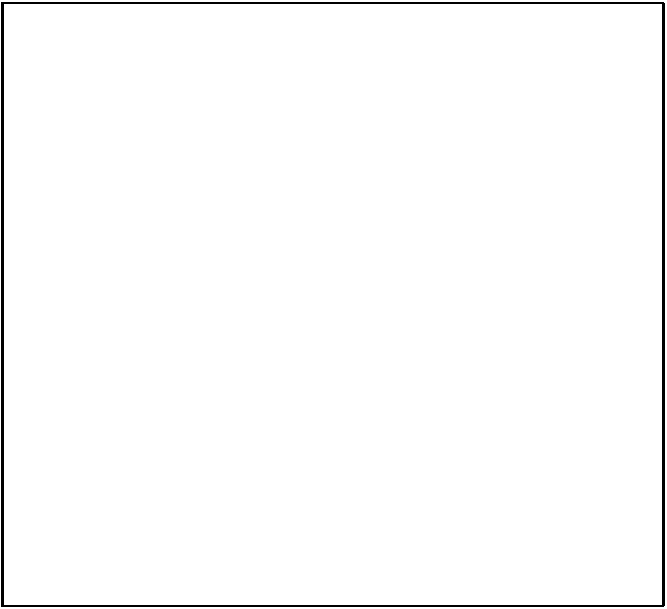

\picplace{8cm}
\caption{Value of the signal-to-noise ratio for the ACF method (straight lines)
and for the method of coaddition of images (lines proportional to the
square-root), as a function of the total
number of galaxies on the field $N_{\rm g}$.
The signal-to-noise of individual galaxies $\sigma_0$ corresponding to
the different line styles are indicated.}
\end{figure}

\subsubsection{Effects of noise on the shear estimate}
Because the gravitational shear is determined from the anisotropy of
the ACF, the sources of noise on the ACF previously discussed affect
the estimation of the shear.  In particular, in the case of a single
object, as for the ACF, the noise degrades the isophotes and randomly
polarize the object.  Consequently, in order to detect a significant
non-zero shear value, the $S/N$ ratio of the ACF must be sufficiently
large, or, in other words, for a given value of $S/N$, only shear
values greater than some threshold $\gamma_0$ can be detected with
sufficient statistical significance.

Figure 3 is a result of a series of Monte-Carlo experiments where for
each of them we measured the mean and the dispersion of the shear of
100 ACFs. Each ACF has a gaussian shape of $7$ pixels diameter. Its
true ellipticity $\gamma_{\rm true}$ is known, and the orientation is
zero (i.e., we choose $\gamma_{\rm true}$ to be real).  A noise
pattern with a given dispersion and mean zero is added on each ACF;
this fixes the $S/N$ of the ACF. In each experiment, the 100 ACFs
differ only in their noise pattern. We have performed 80 experiments,
with 10 values of $S/N$, and 8 values of $\gamma_{\rm true}$.  The
ellipticities are measured using the spatial filter (37) with a
diameter $s=6$ pixels. We also define the dispersion:
\EQ
\sigma_\gamma=\sqrt{\left\langle 
\left[ \left(\gamma_{\rm obs 1}-\gamma_{\rm true}\right)^2
+\left(\gamma_{\rm obs 2}\right)^2\right]\right\rangle}
\EN
with $\gamma_{\rm obs}=\gamma_{\rm obs 1}+{\rm i}\gamma_{\rm obs
2}=\left|\gamma_{\rm obs}\right|e^{2{\rm i}\varphi}$.  At a given
value of the $S/N$, the threshold $\gamma_0$ is defined as a 1-sigma
detection of the ellipticity:
\EQ
\gamma_0=\sigma_\gamma \;.
\EN

Figure 3a shows the relation between $\gamma_{\rm obs}$, the observed
ellipticity value of the ACF and $\gamma_{\rm true}$, its true value.
The apparent discrepancy is an effect of the spatial filter, and will
be corrected  with the calibration of the shear (see section 4.1).
The quantitative dependence of $\gamma_0$ on the $S/N$ of the ACF is
plotted in Figure 3b.

\begin{figure}
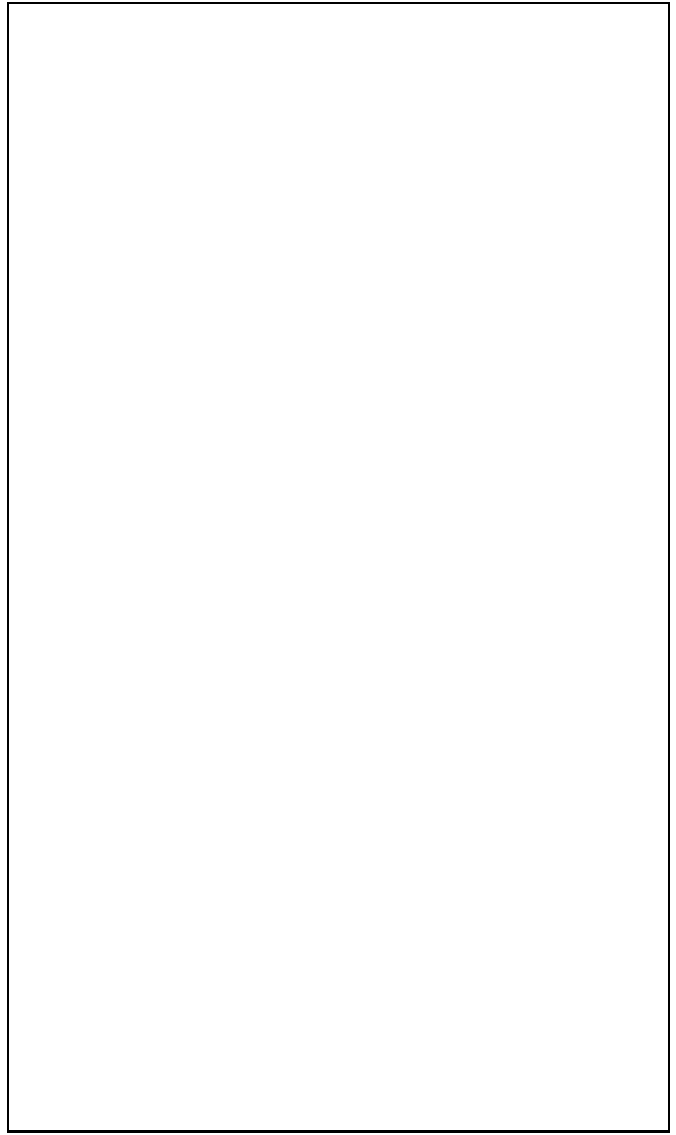

\picplace{15cm}
\caption{The upper plot is the relation between $\gamma_{\rm obs}$ and
$\gamma_{\rm true}$. The proportionality constant is about a factor $4$,
and it is caused by the spatial filter (37) present in the second moment
matrix (26). The lower plot is the 1-sigma detection level of the shear
as a function of the $S/N$ of the ACF.}
\end{figure}

\subsection{ Other sources of errors}
\subsubsection{Errors from the observational characteristics of the galaxies}
The ACF method is ideal if the galaxies have the same flux, the same
size, and the same profile. Unfortunately, this is not the case. A
distribution in size, magnitude, and profile overweight some galaxies,
and can change the {\it statistical properties} of the ACF. In
principle, this could change the value of the shear.
 The detailed analysis of these effects, and how they could
lead to new constraints on the faint galaxies properties, is a scope of the
forthcoming PaperII. Nevertheless we can already 
propose  a solution to the flux-weighting effect.

If the galaxies are distributed in flux, they would (if they had equal
profile and equal size) contribute to the ACF in proportion to the
square of their flux; hence, the brightest galaxies could dominate the
resulting ACF. However, this could be partly avoided in the following
way: define $\gamma(\vc\theta_0;m>m_j)$ as the complex shear as
determined from the ACF, which in turn has been determined by cutting
out all objects on the frame brighter than $m_j$, as described in
Sect.3.1.1. Selecting a set of flux thresholds $m_j$, and a set of
weighting factors $\nu_j$, $1\le j\le M$,  one can form the average 
\EQ
\left\langle\gamma(\vc\theta_0)\right\rangle
={ \sum_{j=1}^M \nu_j\,\gamma(\vc\theta_0;m>m_j)
\over \sum_{j=1}^M \nu_j } \;,
\EN
which is an unbiaised estimate of the shear at $\vc\theta_0$ for all
choices of $\nu_j$. Note that by an appropriate choice of the weight
factors, the contribution by relatively bright galaxies can be suppressed.
One can now try to choose the weight factors in
such as way as to minimize the dispersion of the resulting estimate;
this will be the topic of a future work.

A size and a profile distribution are expected to destroy
the perfectly concentric shapes of the isophotes in the ACF. However,
with the typical smoothing length we used (a Gaussian function with 0.8-1.5
arcmin radius), 
the galaxy number is sufficiently high on a
3-hours exposure at CFHT to make this effect negligible and to induce
an almost noise-free and  regular isophote pattern on the ACF.

\subsubsection{Errors from the sources polarization distribution}
The intrinsic ellipticities of galaxies induce statistical
fluctuations on their mean ellipticity, i.e., on the shear estimated
from that mean. If $\sigma_\epsilon$ is the dispersion of the intrinsic
ellipticity distribution [assumed, as always, that this is isotropic,
so that the mean of the (complex) ellipticity vanishes], then this
intrinsic ellipticity distribution leads to a dispersion $\sigma_{\rm
stat}$ of the resulting shear estimate of
\EQ
\sigma_{\rm stat}={\sigma_\epsilon \over \sqrt{N_{\rm g}}}\; .
\EN
Thus, if the shear is determined from the ellipticity of individual
galaxy images, this provides a limit on the possible accuracy of this
measurement. The dispersion of the intrinsic ellipticity is easily
determined from the data, since it nearly equals the dispersion of the
observed galaxy images. 

An intrinsic ellipticity distribution will also introduce a dispersion
on the shear as estimated from the ACF; however, in this case the
corresponding dispersion can be less securely estimated from the data
itself. However, since the ACF method takes into account information
from much fainter galaxies than can be detected and measured
individually, we can take $\sigma_{\rm stat}$ as an upper bound on the
dispersion introduced by the intrinsic ellipticity distribution.

\subsubsection{Error from instrumental effects}

The PSF is never circular, and an anisotropic correction of the shear
estimate must be applied. The principal sources of anisotropies are
known (see BM). It is possible to measure the
anisotropy from stellar profiles and to model it by an
instrumental deformation matrix $A_{\rm inst}$ as the gravitational
deformation matrix introduced in Section 2.1, Eq.(2). $A_{\rm inst}$
does not contain the PSF, but only the instrumental deformation.
By applying the inverse
instrumental deformation matrix on the shape matrix of the ACF, 
this deformation can be approximately corrected, as in Eq.(9) of Section
2.1:
\EQ
{\cal M}_{\rm true}=A_{\rm inst}^{-1}{\cal M}_{\rm obs}{
A_{\rm inst}^{-1}}^t \;.
\EN
As mentioned by BM, this approximation works provided
that the instrumental deformation is constant in direction, and
does not exceed 5-10 $\%$.

\subsection{Choice of the local smoothing length}
 
The ACF is computed in a given smoothing window of the image where the
distortion is supposed to be approximately constant. The
characteristic size of this area is called the local smoothing
length. Its typical scale must be optimized according to the limiting
shear value and the typical spatial resolution we want to achieve.
First, the number of galaxies $N_{\rm g}$ must be large enough for the
statistical fluctuations of the resulting shear not to exceed the
desired accuracy of the shear measurement.  That is, $\sigma_{\rm
stat}$ of Eq.(48) must be negligable when compared to $\vert
\gamma_{\rm obs}\vert $. Second, $\vert \gamma_{\rm obs}\vert $ must
to be larger than $\gamma_0$ (see Fig.3b), which means that the $S/N$
of the ACF must be sufficiently high. Since it is related by Eq. (44)
to the number of galaxies, $N_{\rm g}$ is also constrained. The choice
of the smoothing length is a compromise between these considerations.
If it is too large, the small scales of the shear are averaged out. If
it is too small, the noise and the statistical fluctuations induce a
false value for the shear.
Table 1 summarizes some constraints on
$N_{\rm g}$ in the light of Sections 3.2.1 and 3.2.2. These values are only
indicative since an exact model for the faint galaxies (size, profile,
magnitude) is needed to make realistic predictions (see paper II).
Such a realistic model will modify the law (44) and changes $N_{\rm g}$.

\begin{table}
\caption[]{Lower limits on $N_{\rm g}$ required to 
measure a shear $\gamma_0$ in a
superpixel of size $256\times 256$ with faint galaxies.}
\begin{flushleft}
\begin{tabular}{lllllllllllllll}
\noalign{\smallskip}
\hline
\noalign{\smallskip}
$\gamma_0$ &0.15&0.08&0.05&0.01\cr
$N_{\rm n}(\sigma_0=3.5)$ &11&21&27&139\cr
$N_{\rm n}(\sigma_0=2.5)$ &22&41&54&274\cr
$N_{\rm n}(\sigma_0=2.0)$ &34&64&86&429\cr
$N_{\rm n}(\sigma_0=1.5)$ &61&114&152&762\cr
$N_{\rm n}(\sigma_0=1.0)$ &137&257&342&1714\cr
$N_{\rm stat}$ &4&14&36&900\cr
\noalign{\smallskip}
\hline
\end{tabular}
\end{flushleft}
\end{table}

$N_{\rm n}(\sigma_0)$ is the minimum number of galaxies with an
individual $S/N$ of $\sigma_0$ required in an area of $256\times256$
pixels (square-shaped top-hat window) to have a $S/N$ of the ACF
sufficiently large to measure a shear $\gamma_0$.  $N_{\rm stat}$ is
the minimum number of galaxies required such that $\gamma_0$ is three
times larger than $\sigma_{\rm stat}$, assuming an intrinsic
ellipticity dispersion of $\sigma_\epsilon =0.1$ [see Eq.(48)].

Moreover, the choice of the smoothing length could depend on the
position on the image since the shear varies spatially. Practically,
an iterative procedure for the choice of this length is hampered by
the very large computational time needed. An easier way is to find the
mean ``best" smoothing length from different tests. However, it should
also be pointed out that for the reconstruction of the surface mass
density of a cluster from image distortions (e.g., Kaiser \& Squires
1993; Fort \& Mellier 1994; Seitz \& Schneider 1995, and references
therein) a spatially homogeneous dispersion on the observed shear
appears to be optimal, i.e., as long as the shear estimate is
unbiased, these reconstructions do not require that a shear different
from zero is measured in each smoothing window.

\section{Applications to data}
Though the ACF method seems theoretically attractive, its 
efficiency has to be demonstrated on real CCD observations. In this
section, we first  show  that the ACF method works on realistic simulated CCD
images where a shear pattern has been introduced. In a second step, we 
analyse the shear field on real data 
corresponding to the two fields already studied by Bonnet et al. (1993,
1994). 

\subsection{Simulated images}
We have used simulated images similar to those shown in BM and where 
photon noises, CCD defects, 
optical and atmospheric artifacts, or flat field residuals can be 
introduced.  The ACF method is tested both on an image containing a
lens, and on an image with no lens. Thus we can check that the method
does not introduce spurious shear, and that the simulations provide an
accurate calibration of the shear as well.  Figure 4 shows the shear
pattern as it is directly measured on the simulated image containing a
lens. The smoothing length is a top-hat filter of size $50''$
which corresponds to about 30 galaxies detectable in each smoothing window.
The ellipticity distribution is a flat distribution between 0.6 and
0.8.
The resulting shear profile, azimuthally averaged with respect to the
lens center, is displayed on Figure 5a.
The errors bars are dominated
by the statistical fluctuations of the galaxy ellipticities;
they are determined from Eq.(48) using $N_{\rm g}\simeq30$. At the
bottom, the 3 measured points before the calibration (49) are plotted 
(see subsection 3.2.2).  To
first order (weak shear approximation), the calibration is a
multiplicative constant and does not depend on the value of the shear 
and the lens model. However, it does depends strongly on the PSF and
the spatial filter used to compute the second moments of the ACF. 
The calibration is found from simulated sheared images which reproduce real
image conditions on telescopes.
The constant is  the ratio of the 
observed (uncorrected) shear  to the true shear.
In Figure 5b, the shear profile
after calibration of the unlensed image is shown. One sees that it is
compatible with zero shear.

\begin{figure}
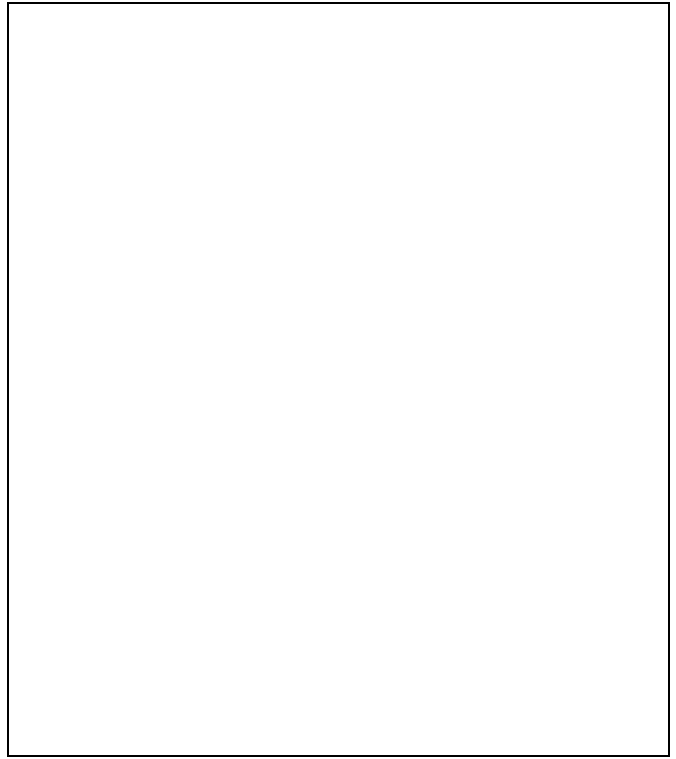

\picplace{10cm}
\caption{Simulation of a 4 hours exposure at CFHT in the B band on a
$3\arcminf5\times 3\arcminf5$ field. The seeing is $0\arcsecf7$ with
no tracking errors. Galaxies are lensed by an isothermal sphere
($\sigma=1000$km/s), with a core radius of $4''$ located at $200''$
left from the field center. The lens redshift is 0.17 and the mean
redshift of the sources is 1. The segments show the local orientation
of the shear. Their length is proportional to the shear intensity.}
\end{figure}

\begin{figure}
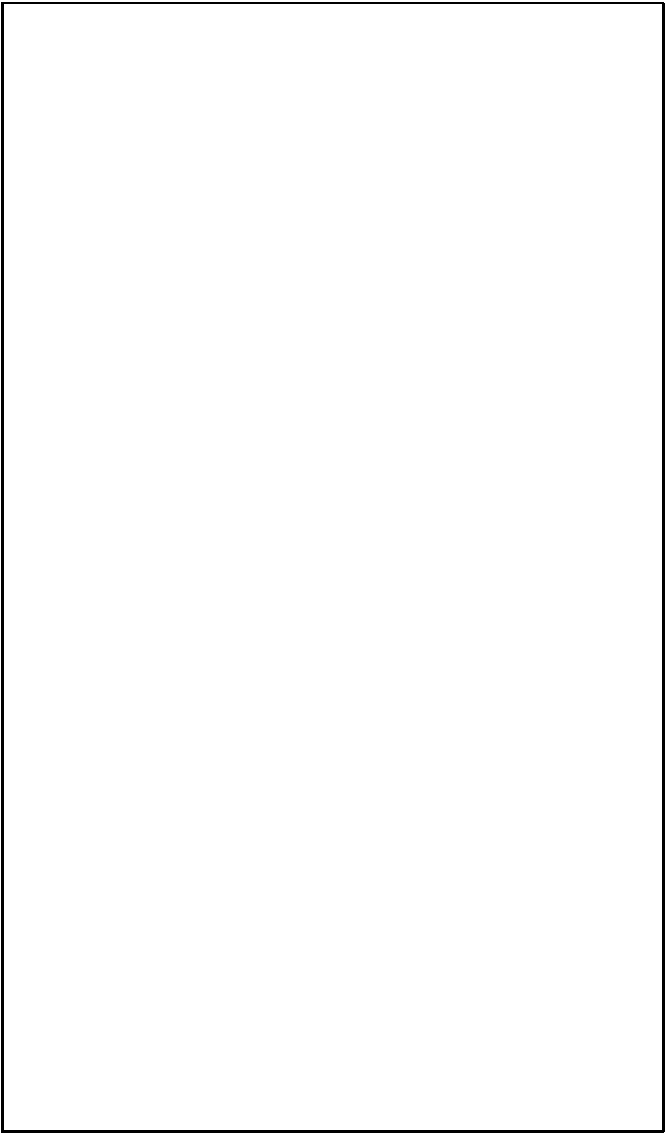

\picplace{15cm}
\caption{(a) 1-dimensional shear profile from the simulation of Fig.
4. At the bottom the uncalibrated measure points are drawn. The
theoretically expected shear profile is plotted as the dashed line. 
(b) The calibrated shear as measured from  the  simulation without a
lens. The measured values are compatible with zero shear}
\end{figure}

\subsection{Real data}
\subsubsection{The shear around Q2345}
The observation of a weak gravitational shear in the field of the
double quasars Q2345+007A,B was made by Bonnet et al. (1993) from deep
CCD images obtained at CFHT. This observational evidence of the
gravitational origin of this doubly imaged quasar was confirmed later
by Mellier et al. (1994), Fisher et al. (1994) and Pell\'o et
al. (1996) who found an excess of galaxies centered on the shear
pattern, and also close to the quasars. They revealed a fairly complex
lens, probably composed of a distant cluster of galaxies and some
bright nearby galaxies.

We analyse here the B and I images of Q2345 used in Bonnet et
al. (1993), and we refer to this paper for the observational details
on the fields. We discard the I-band image for the shear analysis
because the ACF is dominated by a crux pattern (see Figure 1) whose
origin is unclear, but probably associated with residuals from the
addition of many shift and add images which has been moved only along
the pixel directions of the CCDs. Some residuals could be due also to
bad charge efficiency. These defects are damaging for the shear
analysis but they also reveal the high sensitivity of the ACF method
(this suggests that the ACF could be also useful for testing CCDs).

Figure 6a shows the shear pattern found from a set of magnitude
selected objects which clearly confirms the pattern found by Bonnet et
al. (1993). 89 objects are detected in the blue magnitude range
[26,27].  They are surrounded by a circle of size $30$ pixels, whereas
the rest of the image is put to zero. Since the galaxies are detected
here using magnitude criteria, we adopt a completely different
strategy than outlined before. The ACF is computed only from these
selected objects since we detect galaxies here.  The consequence is a
considerable increase of the $S/N$ of the ACF, compared to the case
that the rest of the image was not put to zero.  This is an
interesting alternative of the coadding image in BM, since in the
present approach, the centroid of the individual objects need not to
be determined.  However, the ultra faint objects are lost with this
strategy.  The objects are spatially ordered into 8 groups of $10-15$
members, and the ACF is computed in each of these groups. Since the
signal is highly significant ($S/N \sim 40$), and the shear found is
high (between 10 and 15 $\%$), the measurement error is dominated by
the intrinsic ellipticities of the galaxies. The probability that each
segment (independently of the others) comes from a random orientation
of the galaxies (the ellipticity dispersion is $\sigma_\epsilon=0.2$)
is around $10 \%$. As in the Bonnet et al.  paper, the shear is not
corrected for tracking errors and PSF convolution since there are not
enough stars to evaluate these effects (initially these observations
were not done for measuring weak lensing!).

Figure 6b shows the shear pattern computed from the objects fainter
than $m_B=26.5$. The brighter galaxies are masked, as described in
Sect.3.1.1, and the ACF is computed in a gaussian window of radius
$0\arcminf9$. Since there are many more objects compared to Fig.6a,
the $S/N$ of the ACF is also high ($\sim 40$), and the shear around
the expected excess of galaxies responsible for the image splitting of
the QSO is recovered.  At least $30$ objects are detectable in each
smoothing window, which gives a lower limit of the confidence level
per segment of $99 \%$, assuming the same ellipticity distribution for
the ultra-faint galaxy population as for the ones with $m_B=[26,27]$.
 Remarkably, another shear pattern is detected with the same confidence
level at the bottom of the image. Its superimposition with the I-band
CCD image (Figure 6c) clearly reaveals that the second shear pattern
is centered on a visible excess of galaxies at the top-right of the I
image. At first glance, the shape of this condensation is roughly
elliptical and shows two density peaks.  From the number of galaxies
we see, it is probably a rather rich cluster of galaxies which has
been undetected previously.

\begin{figure*}
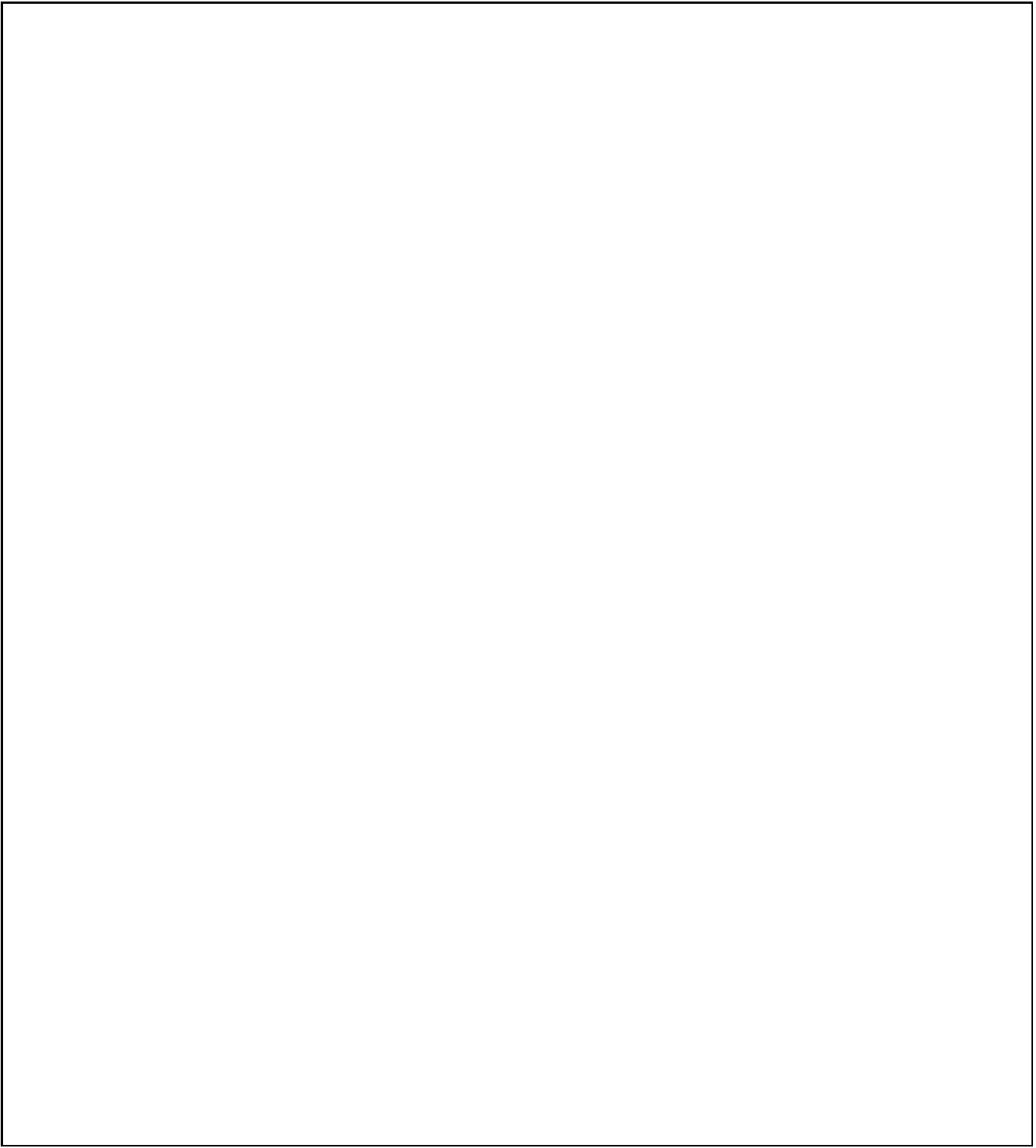

\picplace{20cm}
\caption{Shear patterns found with the ACF method in the field of Q2345+007.
(a) Shear pattern from 89 galaxies with $m_B$=[26,27]. 
(b) Shear pattern from the galaxies with $m_B>$26.5. (c) The shear
pattern from panel (b) overlaid onto the I-band image}
\end{figure*}

\subsubsection{The shear around Cl0024}
The central region of the cluster Cl0024+16 and an external region
located at $6'$ from the cluster center were analysed by 
 BMF, also from deep images 
obtained at CFHT. Again, we refer to this paper
for details about the observations.

The shear analysis done with the ACF of the central part of CL0024 is
displayed in Figure 7a. The Gaussian smoothing length is $40''$, and
the galaxies used are fainter than $m_B=26.5$. The
shape does not show differences with the BMF analysis but the
coherence of the signal around the cluster center is
stronger. The inner pattern clearly shows an
elliptical shape oriented along the second bisecting line, indicating 
an elongated shape for the cluster along this direction. Furthermore,
the innermost central region shows a significant signal in the direction 
perpendicular to the second bisecting line. This signal was not reported
by BMF, probably because of the lower signal to noise ratio as
compared with the ACF. Along with the general orientation reported
above, the central shear resembles what is it expected for a bimodal
mass distribution.

The shear analysis of the large external region is more
interesting. The ACF method was used with 3 differents magnitudes
cuts, $m_V>24, m_V>25$, and $m_V>26$. The corresponding shear maps are
plotted in Figures 7b--d. For each segment, the smoothing
length was $100''$. The general pattern found by BMF is also recovered
here. The brightest map ($m_V>24$) is really similar, and the probable
perturbation also seems to be detected, 
at the same position, though with a
lower significance level.  The intermediate magnitude map looks similar
to the former one, but the perturbation is no more visible. On the
other hand, a significant signal appears at the top right corner (the
northern side of the cluster). It could be an effect of a second
clump. This signal increases significantly on the faintest map, and
there is no doubt at all that it is real. Once we go further away from
the cluster center, the signal become noisier than on previous maps,
but on the other hand, close to the cluster center the intensity of
the shear is even higher than on the brightest and intermediate
magnitude shear map. It suggests that we may be less polluted by
cluster galaxies at such faint magnitudes.

\begin{figure*}
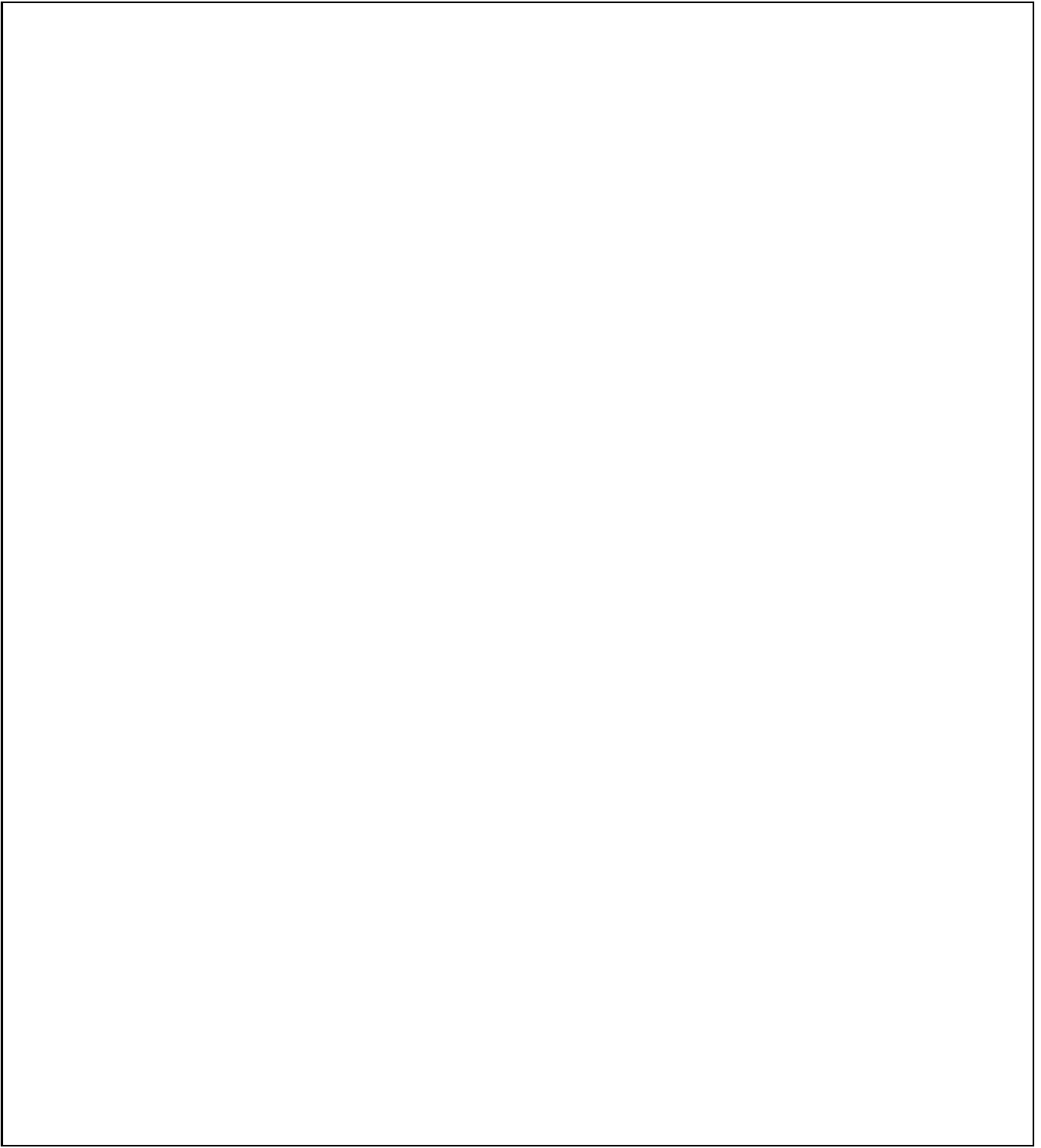

\picplace{20cm}
\caption{Shear maps of Cl0024+1654 obtained with the ACF. (a) The 
central region  of Cl0024+1654 with objects fainter than $m_B=26.5$.
The segments
give the orientation and the intensity of the shear. The coherence is
remarkable. Note also the inner shear pattern which seems oriented along
the second bisecting line. The two large vertical  segments at the bottom
center of the image are artifacts generated after the merging of two CCD
images which overlapped at this position. 
(b) External part Cl0024 with objects fainter than $m_V=$24. 
The overall shape is similar as in BMF. Note the faint signal close to the
perturbation reported by BMF which confirms that we probably
detected a clump of matter toward this direction. 
 (c) Same as (b) for galaxies with $m_V>$25. The main difference to
the previous one is the detection of a clear signal at the top right. 
 (d) Same as (b) for galaxies with $m_V>$26. The shear field at the
top right is also significant and shows that another clump of matter
must be located outside the field of view northward from the cluster
center.}
\end{figure*}

\section{Conclusion}
This paper presents a new approach to probe weakly distorted galaxies
down to the background noise limit. Using the auto-correlation of the
image enables us to increase the signal-to-noise ratio of the shear
considerably, mainly for two reasons: (1) There is no more need to
detect individual objects and to define individual centroids and
individual shape parameters for these galaxies. (2) Even the faintest
galaxies which cannot be detected by standard algorithms provide
information to the ACF. Assuming there is a very dense population of
distant background sources, they improve considerably the
signal-to-noise ratio.  A direct consequence of this correlation
method is its ability to give upper bounds on these ultra-weak
galaxies' number counts (see Paper II).

We have demonstrated that the ACF can be analysed as a single object
whose geometry is related to the ACF of background sources modified by
the gravitational shear. The second moments of the ACF computed over
the whole field provides the magnification map, provided the ACF of
the sources is known. Since the intrinsic ACF is due to very many
faint galaxies, one can hope that it is a universal function, i.e.,
the same in each direction on the sky, so that it can be obtained from
very deep HST images. The relation between the observed shear and the
true shear inferred from the ACF can be determined (e.g., by
simulations), and the intrumental/atmospheric degradations can be
corrected in a standard way, provided the PSF is known, but with a
strong benefit of the high signal-to-noise ratio of the ACF with
respect to individual galaxies.  In our approach we have implicitly
assumed that this population is at high redshift, and does not
correspond to low-redshift galaxies with low surface brightness.

The applications on simulated and real data are very encouraging. The
ACF method easily recovers the shear profile found by BMF with another
and independent method.  However, as shown in the I-band image of
Q2345+007, the drawback is the need of a very high quality of image
aquisition. In particular, the shift and add method must be as random
as possible, preferentialy with no shift in columns and rows, to avoid
privileged directions.

It is also possible to use the ACF method using only selected
galaxies. The practical method is to select the objects with a given
selection criteria (magnitude, or colour) and shade everything which
is outside of a circle centered on each object. The ACF restricted to
these objects can then be computed and analysed as was demonstrated
here. This method was mentioned earlier by Kaiser (1992), though he
did not go into further developments about its practical
implementation.

We are aware that the ACF method is sensible to a flux, size, and
profile distribution of the galaxies.  Although we proposed a solution
for the flux problem, the size and profile problem are more difficult
and will be adressed later.

About the future developments, two main points must be kept in mind.
First, the deep HST empty fields will help in the determination of the
intrinsic correlation function. This leads to the simultaneous
determination of both the magnification and the shear using the
$\chi^2$ minimisation (Eq. 33). However the question whether these
empty fields are really empty is still open, though given that a
one-dimensional function should be determined from observations (i.e.,
the radial profile of the isotropic intrinsic ACF), this does seem
relatively easily possible.  Second, the advantage of the ACF to avoid
the reduction of data process (galaxy detection, first and second
moment estimation) may be increased if a reliable deconvolution of the
PSF on the ACF is possible. However, the proper difficulties of the
deconvolution algorithms, in particular the uncertainties in the
choice of the number of steps for the deconvolution process is still a
source of problems.

The linear dependence of the $S/N$ on the number of galaxies is a
strong argument in favour of the ACF method for the analysis of large
fields where probably a small shear, but coherent over large angular
scales, exists (e.g., Blandford et al. 1991, Kaiser 1992, Villumsen
1996, and references therein). The large number of ultra-faint
galaxies is well adapted to the measurement of small shear.

Another question which should be investigated in future work is how
the shear determined from the ACF method is combined with shear
measurements with the `standard method', i.e., from obtaining the
average ellipticity of observed galaxy images. The combination should
be made in such a way as to minimize the dispersion of the resulting
shear estimate.

\paragraph{Acknowledgements}
We thank H. Bonnet, P.-Y. Longaretti, C. Seitz and S. Seitz for stimulating  
discussions, S. Roques and P. Marechal for enlighting discussions about
the correlation function and M. Dantel-Fort, 
J. B\'ezecourt and J.-M. Miralles  for their helps during the data
reduction. 
This work was partly supported by the ``Sonderforschungsbereich 375-95 f\"ur
Astro--Teil\-chen\-phy\-sik" der Deutschen
For\-schungs\-ge\-mein\-schaft (PS), l'Institut National de Sciences de
l'Univers (INSU)  and the Groupe de Recherche Cosmologie.

\section{References}
\parindent 0truemm 

Bartelmann, M., Narayan, R., 1995, ApJ, 451, 60.

Blandford, R. D., Saust, A., Brainerd, T.,  Villumsen, J., 1991, MNRAS, 251, 600.

Bonnet, H., Fort, B., Kneib, J.-P., Mellier, Y., Soucail, G., 1993,
A\&A, 280, L7. 

Bonnet, H., Mellier, Y., Fort, B., 1994, ApJ 427, L83 (BMF).

Bonnet, H., Mellier, Y., 1995, A\&A, 303, 331 (BM).

Brainerd, T., Blandford, R. D., Smail I., 1996, ApJ, in press.

Broadhurst, T., Taylor, N., Peacock, J. A., 1995, ApJ, 438, 49.

Dalcanton, J. J., 1996, SISSA preprint.

Cole, S., Treyer, M.-A., Silk, J., 1992, ApJ 385, 9.

Fahlmann, G. G., Kaiser, N., Squires, G., Woods, D., 1994, ApJ 437, 56.

Fischer, P., Tyson, J. A., Bernstein, G., Guhathakurta, P., 1994, ApJ, 431, L71.

Fort, B., Mellier, Y., 1994, A\&AR, 5, 239.

Kaiser, N., 1992, ApJ, 388, 272.

Kaiser, N., Squires, G., 1993, ApJ, 404, 441.

Kaiser, N., Squires, G., Broadhurst, T., 1995, ApJ, 449, 460.

Mellier, Y., Dantel-Fort, M., Fort, B., Bonnet, H., 1994, A\&A, 286, 701.

Miralda-Escud\'e, J., 1991, ApJ, 380, 1.

Pell\'o, R., Miralles, J.-M., Le Borgne, J.-F., Picat, J.-P., Soucail, Bruzual, G., 1996, submitted.

Schneider, P., Rix, H. W., 1996, ApJ, submitted.

Schneider, P., Seitz, C., 1995, A\&A, 294, 411.

Schneider, P., Ehlers, J., Falco, E. E., 1992, {\it Gravitational
Lenses}, Springer.

Seitz, C., Schneider, P., 1995, A\&A, 297, 287.

Seitz, C., Kneib, J.-P., Schneider, P., Seitz, S. 1996, A\&A, submitted.

Smail, I., Hogg, D., Yan, L., Cohen, J. G., 1995, ApJ, 449, L105.

Squires, G.,  Kaiser, N., Babul, A., Fahlmann, G. G., Woods, D.,
Neumann, D. M., Bohringer, H., 1996, in press.

Van Waerbeke, L., Schneider, P., Mellier, Y., Fort, B., in preparation (Paper II).

Villumsen, J., 1996, MNRAS, in press.

\end{document}

%% file: anda9.tex
\newcommand{\EQ}{\begin{equation}}
\newcommand{\EN}{\end{equation}}
\newcommand{\EQA}{\begin{eqnarray}}
\newcommand{\ENA}{\end{eqnarray}}